\documentclass [12pt]{article}

\begin{document}
\title{\bf Neutron Star Properties Viewed by the ENU Model}
\author{Miroslav S\'{u}ken\'{\i}k and Jozef \v{S}ima \\[1ex]
  Slovak Technical University, FCHPT, Radlinsk\'{e}ho 9, \\
  812 37 Bratislava, Slovakia
  \\
  e-mail:  sukenik@minv.sk, sima@chtf.stuba.sk} \date{}
\maketitle
\begin{abstract}
\label{sec:abstractne}

Up-to-now known characteristics of radio pulsars, such as mass limits, 
magnetic field intensity, rotational period, and maximum radiation are 
mainly of empirical nature. Applying the Expansive Nondecelerative Universe 
model (ENU) into the issue allows to offer a deeper theoretical explanation 
of the known parameters and to estimate their limits. Using the ENU approach 
the following values related to synchrotron radiation emitting radio pulsars 
were estimated: the lower and upper limits of magnetic field intensity are 
$B_{P(\min )} \cong 8.5\times 10^{6}$\,T, and $B_{P(\max )} \cong 4.4\times 
10^{9}$\,T, respectively, the maximum rotation period reaches 3.9\,s, the 
maximum radiation output of a pulsar is $P_{P(\mathit{rad},\max )} \cong 5.6\times 
10^{29}$\,W (all the values relate to radio pulsars with 1.4 solar masses and 
radius $r = 10^{4}$\,m and are mass and radius dependent). These values are 
in accordance with the experimental observations.


\end{abstract}

\section{INTRODUCTION}

Neutron stars are objects with the mass from about 1.4 (the
Chandrasekhar limit) to 3.5 solar masses, diameter $r \cong
10^{4}$\,m, and density $10^{17}$\,kg\,m$^{{\rm -} {\rm 3}}$. Spinning
neutron stars emitting electromagnetic radiation from their poles are
called pulsars. The radiation energy varies from radio waves to gamma
rays. Until now investigated neutron stars are characterized [1 - 3]
by the magnetic field ranging from $10^{7}$\,T to $10^{9}$\,T and the
rotation period from miliseconds to seconds.  Rotating magnetic field
influences the electrons being present in the environment, giving thus
rise the formation of synchrotron radiation. In addition to
"classical" radio pulsars, soft gamma repeaters - magnetars - with
magnetic fields up to $10^{11}$\,T are known, their radiation is not,
however, of synchrotron nature caused by electrons. Magnetars are
outside the scope of this article.

Radiation output of radio pulsars reaches usually $10^{24} - 10^{28}$\,W, 
with $10^{30}$\,W as a known maximum. The extremely strong gravitational 
field is able to attract and hold neutrons and electrons within the spheres 
above iron surface crust up about 1 cm or 10 m, respectively.

Owing to Vaidya metric application [4], the model of Expansive 
Nondecelerative Universe [5] enables to localize gravitational energy [6]. 
Stemming from a general formula [6], the absolute value of gravitational 
energy density $\varepsilon _{g} $ at a pulsar surface can be expressed as
\begin{equation}
\label{eq:1a}
{\left| {\varepsilon _{g}}  \right|} = {\frac{{R c^{4}}}{{8\pi G}}} 
\cong {\frac{{3m_{P} c^{2}}}{{4\pi  a  r_{P}^{2}} }} \cong 4.6\times 
10^{12} \mbox{\,J/m$^3$}
\end{equation}
where $R$ is the scalar curvature ($R \ne 0$ in Vaidya metric [4, 6]), 
$m_{P} $ is the pulsar mass, $r_{P} $ is its radius, and $a$ represents the 
gauge factor (in the above and following equations the mass $2.8\times 
10^{30}$\,kg, radius $10^{4}$\,m, and gauge factor $1.3\times 10^{26}$\,m were 
introduced).

It can hardly be a coincidence that the gravitational energy density is very 
close to (just about 1.4 times higher than) the electromagnetic energy 
density of hydrogen atom is. 

Gravitational field may be described by a wave function [6] 
\begin{equation}
\label{eq1}
\Psi _{g} = \exp ( - i\omega _{g} t)
\end{equation}
where $\omega _{g} $ is the frequency of gravitational wave.

At the pulsar surface
\begin{equation}
\label{eq:3a}
\omega _{g} = \left( {{\frac{{m_{P}  c^{5}}}{{\hbar  a  r_{P}^{2}} }}} 
\right)^{1 / 4} \cong 1.5\times 10^{18} \mbox{\,Hz}
\end{equation}
Based on the fact that the gravitational field of a pulsar is able to held 
electrons up to 10 m distance from the surface it follows that the magnetic 
moment vector of the electrons shall perform precessional motion with the 
frequency
\begin{equation}
\label{eq2}
\omega _{e} = {\frac{{B  \mu _{e}} }{{\hbar} }} = {\frac{{B  e}}{{m_{e} 
}}}
\end{equation}
where $B$ is the pulsar magnetic field intensity, $\mu _{e} $ is the 
electron magnetic moment, $m_{e} $ and $e$ are the electron mass and charge, 
respectively.

The pulsar stability is preserved only when
\begin{equation}
\label{eq3}
\omega _{g} \le \omega _{e} 
\end{equation}
In case of equality (\ref{eq3}), stemming from (\ref{eq:3a}) to
(\ref{eq3}) the lower limit of pulsar magnetic field intensity follows
as
\begin{equation}
\label{eq:6a}
 B_{P(\min )} \cong 8.5\times 10^{6} \mbox{\,T}
\end{equation}
which is in excellent accord with the value obtained from experimental 
observations. The upper limit of pulsar magnetic field intensity can be 
estimated based on the Compton frequency of electron
\begin{equation}
\label{eq:7a}
 \omega _{C} = {\frac{{m_{e} c^{2}}}{{\hbar} }} \approx 10^{21} \mbox{\,Hz}
\end{equation}
where the limiting condition
\begin{equation}
\label{eq4}
\omega _{e} = \omega _{C} 
\end{equation}
\begin{equation}
\label{eq:9a}
B_{P(\max )} \cong 4.4\times 10^{9} \mbox{\,T}
\end{equation}
Of course, the frequency $\omega _{e} $ can approach but never reach the 
value of $\omega_{C} $. 

As to the structure and composition of neutron stars, various
hypotheses have been formulated (from iron-like crust to quark-gluon
plasmas). Further we show another mode to derive the value of
$B_{P(\max )}$. Suppose, whole pulsar consists of particle with the
mass of electron. A number of electrons $n(e)$ corresponding to a
pulsar of the mass $m_{P} $ is then given as
\begin{equation}
\label{eq5}
n(e) = {\frac{{m_{P}} }{{m_{e}} }}
\end{equation}
In such a case it can be supposed that the maximum rotation energy of the 
pulsar is
\begin{equation}
\label{eq6}
E_{P(\mathit{rot},\max )} = \frac{m_P \hbar  \omega_{P(\max )}}{m_e} = 
\frac{m_P \hbar  e  B_{P(\max)} }{m_e^2} 
\end{equation}
where $\omega _{P(\max )} $ is a maximum procession motion of the
electron magnetic moment vector at the maximum magnetic field
intensity $B_{P(\max )}$. The upper limit of rotational energy of the
spherical bodies is expressed as
\begin{equation}
\label{eq7}
E_{(\mathit{rot},\max )} = {\frac{{m  c^{2}}}{{5}}}
\end{equation}

Putting (\ref{eq6}) and (\ref{eq7}) identical, it leads to
\begin{equation}
\label{eq:13a}
B_{P(\max )} \cong 6.7\times 10^{8} \mbox{\,T}
\end{equation}
which is the value being in good agreement with expectations. It can be 
stated that there is no possibility to find a pulsar of a 1.4 solar masses 
having its magnetic field intensity higher than that given by (9) or (13). 

The lower limit of pulsar rotational energy emerges when the electron mass 
in (\ref{eq6}) is substituted for the neutron mass $m_{n} $ and the minimum value 
of the magnetic field intensity $B_{P(\min )} $ given by (6) is introduced. 
In such a case,
\begin{equation}
\label{eq8}
{\frac{{m_{P} \hbar  e  B_{P(\min )}} }{{m_{n}^{2}} }} = {\frac{{m_{P} 
r_{P}^{2} \omega _{(\min )}^{2}} }{{5}}}
\end{equation}
The maximum rotation period of a neutron star following from (\ref{eq8}) is then
\begin{equation}
\label{eq:15a}
t_{(rot,\max )} = {\frac{{2\pi} }{{\omega _{(\min )}} }} \cong 3.9
\mbox{\,s}
\end{equation}
Also this value is in excellent agreement with observation. At present, the 
maximum rotational period of 3.8 s is reported, it should be pointed out, 
however, that 1) the rotational period is mass and radius dependent, 2) it 
can change due to energy emission, 3) it can change due to mass transfer 
when existing in binaries. Longer rotational periods are usually ascribed to 
white dwarfs.

The ENU approach enable to evaluate the radiation output of pulsars
$P_{P(rad)} $. In order to secure a pulsar stability, its radiation
output cannot exceed its gravitational output $P_{P(g)} $, i.e.
\begin{equation}
\label{eq9}
P_{P(g)} \ge P_{P(rad)} 
\end{equation}
In the ENU model, generally [6]
\begin{equation}
\label{eq10}
{\left| {P_{g}}  \right|} = {\frac{{d}}{{dt}}}\int {{\frac{{R  c^{4}}}{{8  
\pi  G}}}dV = {\frac{{m  c^{3}}}{{a}}}} 
\end{equation}
Comparing eqs. (\ref{eq9}) and (\ref{eq10}) is follows that any pulsar formed from a star 
with the Chandrasekhar limit mass and radius $r \cong 10^{4}$\,m cannot have 
radiation output higher than 
\begin{equation}
\label{eq:18a}
P_{P(\mathit{rad},\max )} \cong 5.6\times 10^{29} \mbox{\,W}
\end{equation}
which corresponds to the observed values.

\subsection{Conclusions}

Up-to-now known values of pulsar magnetic field intensity, rotational 
period, and maximum radiation output stem from experimental observation and 
are of empirical nature. Applying the ENU model into the matter allows to 
offer a deeper theoretical explanation of the known values and to estimate 
the limits for the mentioned parameters.

\subsection{References}

\begin{description}
\item {}[1] B. Link, R.I. Epstein, J.M. Lattimer, Phys. Rev. Lett., 83 (1999) 3362

\item {}[2] I. Bednarek, R. Manka, Int. J. Mod. Phys. D, 10 (2001) 607 

\item {}[3] N.K. Glendenning, Compact Stars: Nuclear Physics, Particle Physics and 
General Relativity (2$^{{\rm n}{\rm d}{\rm} }$ed.), Springer, New York, 2000 

\item {}[4] P.C. Vaidya, Proc. Indian Acad. Sci., A33 (1951) 264 

\item {}[5] V. Skalsk\'{y}, M. S\'{u}ken\'{\i}k, Astrophys. Space Sci., 178 (1991) 
169 \& 181 (1991) 153

\item {}[6] J. \v{S}ima, M. S\'{u}ken\'{\i}k, Spacetime \& Substance 3 (2001) 125
\end{description}

\end{document}